# Journal Maps on the Basis of Scopus Data:

# A comparison with the *Journal Citation Reports* of the ISI

Loet Leydesdorff [a], Félix de Moya-Anegón [b] & Vicente P. Guerrero-Bote [c]

**Abstract**

Using the *Scopus* dataset (1996-2007) a grand matrix of aggregated journal-journal citations was constructed. This matrix can be compared in terms of the network structures with the matrix contained in the *Journal Citation Reports* (*JCR*) of the Institute of Scientific Information (ISI). Since the *Scopus* database contains a larger number of journals and covers also the humanities, one would expect richer maps. However, the matrix is in this case sparser than in the case of the ISI data. This is due to (*i*) the larger number of journals covered by *Scopus* and (*ii*) the historical record of citations older than ten years contained in the ISI database. When the data is highly structured, as in the case of large journals, the maps are comparable, although one may have to vary a threshold (because of the differences in densities). In the case of interdisciplinary journals and journals in the social sciences and humanities, the new database does not add a lot to what is possible with the ISI databases.

[a] Amsterdam School of Communications Research (ASCoR), University of Amsterdam, Kloveniersburgwal 48, 1012 CX Amsterdam, The Netherlands; loet@leydesdorff.net; http://www.leydesdorff.net .
[b] CSIC, CCHS, IPP, SCImago Research Group, Spain; felix.moya@scimago.es.
[c] Universidad de Extremadura, Information and Comunication Science Department, Spain, guerrero@unex.es

## Introduction

Derek de Solla Price's (1961) fascination with the growth of journal literature during more than three centuries was followed by both Garfield and Narin as early pioneers of the bibliometric enterprise. Narin *et al*. (1972) elaborated on Price's (1965) suggestion to use aggregated journal-journal citations for multivariate analysis in order to distinguish groups of journals representing specialties, and Garfield (1972) developed citation analysis as a tool for journal evaluation. Ever since, a large number of indicators for journals have been developed, and science mapping on the basis of aggregated journal-journal citations has become an industry (e.g., Rosvall & Bergstrom, 2008; Klavans & Boyack, 2009; Leydesdorff & Rafols, 2009; De Moya-Anegón *et al*., 2008).

Hitherto, maps and indicators have been largely confined to the domain of the journals included in the *Science Citation Index* (*SCI*) and the *Social Science Citation Index* (*SSCI*) of the Institute of Scientific Information (ISI of Thomson-Reuters). In 2004, both *Scopus* and *Google Scholar* were launched, providing alternatives to the ISI databases for citation analysis. Although one could in principle construct an aggregated journal-journal citation matrix from the *Google* database, this may not be an easy task (Leydesdorff, 1994). Unlike the *SCI* and *Scopus*, *Google Scholar* is not delimited in terms of journals processed at the backend, but on searching the web for scholarly contributions. Journal names in this case are only attributes to documents and are not controlled for variants and misspellings.

Both the *Science Citation Index* and *Scopus* are based on a delineation in terms of scholarly journals. The contents of these so-called "source journals" are processed bibliographically in terms of publications and the references provided by the authors of these publications. These references include citations to "non-source" journals, books, reports, patents, etc. Both Elsevier and the ISI identify a number of journals as additional source journals when they are cited without including them in terms of the citing journals. In the *Scopus* database journals are identified with a unique item identifier which is kept constant over the years, whereas in the (*Social*) *SCI* the titles of the journals (and their corresponding abbreviations) function as identifiers.

The ISI aggregates the aggregated citation data among journals in the yearly *Journal Citation Reports* (*JCR*). *JCRs* are available since 1974 and 1977 for the *Science Citation Index* and the *Social Science Citation Index*, respectively. Electronic versions on CD-Rom have been available for these two databases since 1994; since 1997 the data is also available online. At the time of this project, the latest (2007) version of the two ISI databases contained 7,940 journals. Note that there is no *JCR* available for the *Arts & Humanities Citation Index* (*A&HCI*).

Hitherto, Elsevier has not published the aggregated journal-journal citations among journals contained in the *Scopus* database. In 2007, this database contained 23,503 journals with an item identifier; that is, approximately three times as many as the source journals in the two *JCRs* combined (7,940 in 2007).[1] However, only 14,964 (63.7%) of these journals were processed by *Scopus* in 2007. The ISI processed 7,925 (99.8%) of the source journals. In other words, the *Scopus* references draw upon a much larger pool than the ones included in the *JCR*s.

[1] In 2008, the number of included journals was increased to 23,553.

The *Scopus* (1996-2007) data is made available for scientometric analysis by the *Scimago* group in Spain (at http://www.scimago.es). This data contains the full set provided with an item identifier since 1996 (that is, 23,503 journals). In principle, a set of the cited or citing journals for each year can be derived from this database by using a dedicated routine. We used this set containing aggregated citations among all journals with item identifiers during the period 1996-2007.

The unique identification in the *Scopus* database is limited backwards to 1996, while the ISI database includes all references provided in a certain year. This includes both references to non-source journals and to sources of older dates in the case of the ISI set, while the *Scopus* dataset used in this study contains only references to the source journals identified as items during the period 1996-2007.[2] However, the ISI database allows us additionally to distinguish between the citations to sources in the last ten years (1998-2007) and the rest. We shall test below for the effects of this disaggregation of the ISI database and compare the results with those obtained using *Scopus* data covering the period 1996-2007.

---

[2] Additionally, the *Scopus* database contains a large number of journals which were imported from MedLine (as a public database) or previously organized databases of Elsevier such as Embase and Compendex. The record before 1996 contains another 19 million references, but these were not systematically processed on the basis of the selection of source journals (Ove Kahler at *Scopus*, *personal communication*, 27 August 2009).

The *Scopus* data allows us to ask questions different from those addressed to the ISI data. For example, the arts and humanities are included, and the extension may include many more journals in languages other than English. The two databases can also be compared on other bibliometric characteristics such as citation patterns or the *h*-indices of authors (Bar-Ilan, 2008; Jacso, 2005 and 2008; Meho & Rogers, 2008; Meho & Yang, 2007). Focusing on a subset of oncology journals, López-Illescas *et al.* (2008) found that with reference to this set, the ISI database could be considered as a subset of the *Scopus* database. Although the coverage of the latter database was 90% larger, the average impact factor in the latter set was only 2.6% higher than in the ISI set.

Klavans & Boyack (2007) compared maps based on co-citation analysis at the document level in both databases. Klavans & Boyack (2009a) argue that a consensus map of science seems possible based on communalities between the two databases (cf. Rafols & Leydesdorff, 2009). Recently (July 2009), Elsevier launched a facility *SciVal* (at http://info.funding.scival.com/ ) based on a further extension of Klavans & Boyack's (2009b) co-citation maps of the *Scopus* data.

We pursue the analysis at the level of journals. Our original research question was to investigate comparatively the informative quality of journal maps on the basis of data in the two databases from a user perspective. Raising questions about the differences, however, brought us to a (second) analytical question about the different qualities of the data for this purpose. Can the differences also be explained? As noted, a major difference is the cutoff in the *Scopus* data in 1996: data older than 1996 are included, but not systematically marked with item identifiers, and therefore not usable for our purpose without considerable data cleaning. The expectation is that the data of *Scopus* (1996-2007) are richer than the *SCI* (1998-2007) because of the larger coverage, but potentially less informative than the full set of the *SCI* because of the cut-off of the historical tails of the distributions.

The two databases are also different in size. There has been continuous debate about how the ISI selects journals for inclusion as source journals in the database (Garfield, 1979, 1990; Testa, 1997). Sivertsen (2003) found no bias in the coverage of Scandinavian publications by the *Science Citation Index*. However, the ISI concedes that the selection system for inclusion in the *Science Citation Index* may be biased against journals written in non-latin alphabets. Special care has been taken in the past to include Russian and Japanese journals into the database (Garfield, 1979, 1998). *Scopus* brings a document online specifying its criteria for inclusion of journals and other content into the database at http://info.scopus.com/docs/content_coverage.pdf (retrieved on August 7, 2009).

As reported by Thomson-Reuters Scientific in May 2008,[3] the *SCI* was recently extended with another 700 journals in order to improve its "regional" coverage. Leydesdorff & Wagner (2009) found a notable effect of this extension on the coverage of Eastern European contributions, but less so for other world regions. This recent expansion seems to correspond to the larger coverage by the *Scopus* database in these linguistic domains. In other words, the two databases are competing with each other on the market of academic libraries. In June 2009, the *Scopus* team announced to increase the coverage of journals in the arts and humanities to 3,500 by adding all (approximately 1,450) top global journals using the European Reference Index for the Humanities (ERIH) of the European Science Foundation.[4] These recent extensions are not included in our data; the analyses are based on the tape years 2007 for both datasets.

[3] http://isiwebofknowledge.com/products_tools/multidisciplinary/webofscience/contentexp/expansionessay/

## Language coverage of the journals in the two databases

A major advantage of the larger coverage of the *Scopus* data in terms of journals included could be the coverage of journals in languages other than English. *Scopus* advertizes at its website to "cover titles from all geographical regions including non-English titles as long as English abstracts can be provided."[5] As noted, the lack of coverage in the non-English domain has been a major complaint about the ISI databases (Maricic, 1997; Van Leeuwen *et al.*, 2001). This can particularly be considered as a problem for the evaluation of research performance in the social sciences and the humanities (Glänzel & Schöpflin, 1999; Nederhof *et al*., 2001; Nederhof, 2006).

For example, the French CNRS subsidizes approximately 225 journals which are only partially covered by the U.S.-based *Science Citation Indices* (De Looze *et al*., 1996; Legentil, personal communication). Among the 4000+ journals published in Chinese only 17 were included in the ISI database in 2007 (Ren & Rousseau, 2002; Jin *et al*., 2002; Leydesdorff & Jin, 2005; Moed, 2002). Eighty journals in the *JCRs 2007* were published in China (but not necessarily in Chinese). Of the 7,940 journals included in the ISI database, 6,912 (87.1%) are published in English. Another 10.1% of the journals are multilingual, but mostly include English as a main language. Journals in other languages are marginal to the database ( Table 1).

---

[4] "Scopus works with European foundation to expand Arts & Humanities coverage", at http://info.scopus.com/news/press/pr_100609.asp, June 10, 2009.
[5] At http://info.scopus.com/detail/what/ [retrieved on August 9, 2009].

| | *ISI* | *%* | *Scopus 1996-2007* | *%* | *Scopus Aug. 2009* | *%* |
|---|---|---|---|---|---|---|
| English | 6,912 | 87.1 | 14,109 | 81.1 | 14,193 | 80.7 |
| Multi-Language | 802 | 10.1 | 2,993 | 17.2 | 1,227 | 7.6 |
| German | 88 | 1.1 | 44 | 0.3 | 363 | 2.1 |
| French | 45 | 0.6 | 67 | 0.4 | 292 | 1.7 |
| Spanish | 33 | 0.4 | 70 | 0.4 | 360 | 2.1 |
| Chinese | 17 | 0.2 | 0 | 0,0 | 323 | 1.8 |
| Japanese | 13 | 0.2 | 0 | 0.0 | 126 | 0.7 |
| Russian | 13 | 0.2 | 1 | 0.0 | 36 | 0.2 |
| Portuguese | 5 | 0.1 | 19 | 0.1 | 110 | 0.6 |
| Italian | 2 | 0.0 | 23 | 0.1 | 134 | 0.8 |
| Rumanian | 2 | 0.0 | 1 | 0.0 | 7 | 0.0 |
| Croatian | 1 | 0.0 | 4 | 0.0 | 33 | 0.2 |
| Dutch | 1 | 0.0 | 1 | 0.0 | 30 | 0.2 |
| Eskimo | 1 | 0.0 | 0 | 0.0 | 0 | 0.0 |
| Korean | 1 | 0.0 | 0 | 0.0 | 15 | 0.1 |
| Polish | 1 | 0.0 | 21 | 0.1 | 69 | 0.4 |
| Slovene | 1 | 0.0 | 0 | 0.0 | 6 | 0.0 |
| Swedish | 1 | 0.0 | 0 | 0.0 | 6 | 0.0 |
| Turkish | 1 | 0.0 | 13 | 0.1 | 44 | 0.3 |
| Hungarian | 0 | 0 | 10 | 0.1 | 24 | 0.1 |
| Czech | 0 | 0 | 2 | 0.0 | 30 | 0.2 |
| Slovakian | 0 | 0 | 2 | 0.0 | 6 | 0.0 |
| Lituanian | 0 | 0 | 1 | 0.0 | 1 | 0.0 |
| Danish | 0 | 0 | 1 | 0.01 | 1 | 0.0 |
| Iceland | 0 | 0 | 1 | 0.01 | 0 | 0.0 |
| *other* | 0 | 0 | 0 | 0 | 158 | 1.0 |
| Total | 7,940 | | 17,383 | | 17,594 | |

**Table 1**: Distribution of journals in the ISI and *Scopus* databases in terms of languages. The ISI distribution is based on the set in 2007; the *Scopus* database on the journals included from 1996 to 2007.

Table 1 informs us that this promise of a wider coverage of languages by the *Scopus* database can be nuanced. According to Table 1, the *Scopus* 1996-2007 database would not contain Chinese, Japanese, or Korean journals in 2007. Because we aggregated the journal data from the documents, some journals in these Asian languages are included in the multi-language category. The extension in the French and German domain seems less substantial than in some of the other (mainly European) languages. For this reason we checked with Elsevier (Ove Kahler at *Scopus, personal communication*, 28 August 2009) who was so kind to provide us with data for the fifth and sixth column, including the breakdown of the multi-language category as registered by *Scopus* at the journal level. Since the totals are approximately the same, one may also assume that the language distribution in terms of percentages is approximately the same in our data.

## Methods and materials

The aggregated journal-journal citation data of the ISI was harvested from the CD-Rom version of the *Journal Citation Reports* 2007. This data is identical to the online version of the *JCR*. However, we combined the two databases for the *Science Citation Index* and the *Social Science Citation Index* into a single *JCR.* A similar matrix was constructed using the *Scopus* database on the basis of the so-called item identities in both the cited and citing items. Because these item identities are unique, the *Scopus* database does not contain errors. Misspellings of journal names, however, are included in the *SCI* as references to non-source journals. For example, Leydesdorff (2008, Table 4, at p. 285) found 14 so-called “non-source journals” in the *SCI* based on common misspellings of journal names with values above 10,000 in 2005.

Both datasets were brought under the control of a relational database management system and exported to *Pajek*, a freeware program for the visualization and analysis of networks.[6] The resulting files are asymmetrical matrices which can be analyzed in both the cited and the citing dimensions in terms of journal citation patterns. We use SPSS v.15 for the factor analysis and focus on the cited dimension unless stated otherwise. For the comparisons between the three sets (*Scopus* 1996-2007, *SCI* 1998-2007, and the complete *SCI*), we extract ego-networks of seed journals using dedicated software. All journals contributing one percent or more to the being-cited totals of citations of the seed journal in 2007 are included in these ego-networks. We use the same sets of journals when comparing the original ISI data with the truncated set *SCI* (1998-2007) in order to facilitate the comparison and for the technical reason that we have no easy hold in the data for adjusting the chosen thresholds of one percent in the reduced sets. In all cases, the cosine among the citation patterns is used for normalization (Ahlgren *et al*., 2003) with a threshold of cosine > 0.2. We keep the procedures as constant as possible across the different journal comparisons between the databases.

In the visualizations of the cosine matrices (using Pajek) the sizes of the nodes are vertically proportionate to the logarithm of the number of citations, and horizontally to the logarithm of this number after correction for "within journal self-citations." The width of the links is proportionate to the cosine value (above the threshold of cosine > 0.2). Colors were automatically generated using the *k*-core algorithm in Pajek. The factor structures are indicated in the figures using ellipses with dotted lines; factor designations were provided by the analysts. The number of factors is reasoned on the basis of screeplots in the respective analyses.

[6] Pajek is freely available for academic use at http://vlado.fmf.uni-lj.si/pub/networks/pajek/ .

## Descriptive statistics

Table 2 provides some descriptive statistics of the three citation matrices (*Scopus* 1996-2007, *SCI*, and *SCI* 1998-2007). In addition to the citation indicators, we computed betweenness centralization and the average clustering coefficients as network measures because these two indicators teach us about the coherence and the decomposability of the databases, respectively.[7]

| | | *Scopus 2007* (1996-2007) | *JCR 2007* ( - 2007) | *JCR 2007* (1998-2007) |
|---|---|---|---|---|
| a. | Number of source journals | 23,503[8] | 7,940 | 7,900 |
| b. | Number of source journals (citing) | 14,964 | 7,925 | 7,793 |
| c. | unique journal-journal relations among source journals | 3,570,020 | 1,460,847 | 1,318,180 (= 90.23%) |
| d. | Total "cited" | 22,370,409 | 27,477,499 | |
| e. | Total "citing" (source journals) | 28,492,874[9] | 24,979,391 | 16,549,053 |
| g. | **Sum journal-journal relations** | **22,370,409** | **24,979,391** | **16,549,053 (= 66.25%)** |
| h. | Average non-zero cell value ( = g/c) | 6.27 | 17.10 | 12.55 (= 73.39%) |
| i. | Density | 0.0065 | 0.0232 | 0.0209 |
| j. | Largest component | 13,376 (89.4%) | 7,694 (96.9%) | 7,674 (97.1%) |
| k. | Betweenness centralization | 0.05198 | 0.08354 | 0.08628 |
| l. | Average Clustering Coefficient | 0.2700429 ± 0.1719204 | 0.3478451 ± 0.1255273 | 0.3460075 ± 0.1289489 |

**Table 2**: Summary statistics of the three databases under study.

7 Degree centralization could not be computed since the networks contain loops; closeness centralization could not be computed since the networks are not strongly connected.

8 In the case of the *Scopus* data, this is the number of journals with an item identity in the period 1996-2007. Of these journals, 22,238 were processed in the cited direction. Journals may be included in the database in one or more years without necessarily presence during the whole period under study.

9 This number was kindly provided to us by one of the referees. Our analysis of the *Scopus* data aggregates only identified item-item relations without further attempt to resolve the cited journal name to keys which were not already provided in the data.

More detailed analysis of the data revealed that only 14,964 journals of the 23,503 journals with item identifiers included in *Scopus* (1996-2007) are processed as source journals (citing). (Only 13,699 were processed as both cited and citing in 2007.) In the case of the ISI set, the "All others" category leads to discrepancies between "total cited" (d.) and "total citing" (e.). The ISI database also contains a large number (292,105) of non-ISI source journal names among the references, which are in principle available for further analysis. The huge number of non-source items in *Scopus* (2,664,407)[9] was not available for analysis. This study focuses on the 24,979,391 unique journal-journal relations among the 7,940 journals in the case of the ISI and 22,370,409 such relations in the case of *Scopus* (boldfaced as line g. in Table 2).

The descriptive statistics for the *SCI* (1998-2007) subset are added in the third column of Table 2. The aggregated number of references to ISI source publications of 1997 or older is 8,430,338.[10] This amounts to 33.75% of the total of 24,979,391 references included in the analysis. However, 90.23% of the unique journal-journal relations (c.) in the full set are also covered by the reduced set. The density (i.) of the reduced set consequently is not much lower than that of the full set, but the average non-zero cell value (h.) is. This average may be low in the *Scopus* database because of the inclusion of journals with lower citation rates.

The numbers of journal-journal relations are of the same order in both databases, but the number of journals is larger in the *Scopus* database than in the two *Science Citation*

[10] The number of references to the 7,793 journals which contain citations to the last ten years was 24,554,342 of which 8,005,289 are given to the cited volumes older than 1998. The resulting 16,549,053 ( = 24,554,342 - 8,005,289) is provided in Table 2 as the sum of journal-journal relations in this set.

*Indices* combined. Accordingly, the density is larger in the ISI database. In other words, the aggregated journal-journal citation matrix based on the *Scopus* database contains many more sparse areas than the ISI database. One would expect that this difference is caused by the limitation of the *Scopus* database—reaching back only to 1996—and the larger number of journals covered. Cutting off the historical tails of the distributions in the *SCI*, however, does not have a large effect on the density which remains in the *SCI* (1998-2007) higher than in the *Scopus* (1996-2007) database.

One should keep in mind that the matrix of citations among journals in the case of *Scopus* (1996-2007) is based on aggregation of citations among identified documents. While the ISI database covers the source journals cover-to-cover—although it uses only the so-called citable issues (articles, reviews, letters, and proceedings papers) as the denominator of the impact factor—*Scopus*, for example, does not process cited references in book reviews (Ove Kahler at *Scopus*, *personal communication*, 28 August 2009). In other words, the different organization of the *Scopus* database may have a large effect that can be expected to vary across fields of science (Price, 1970).

For example, for the *Science Citation Index* the percentage of citations in the historical tail (1997 or older) is 33.16%, while it is 44.41% for the *Social Science Citation Index*.[11] All network measures are lower in the *Scopus* database than in the ISI database, but for the latter database it makes not so much difference in terms of the values of the network measures whether one includes the historical tails or not. The structural properties of the database are preserved when focusing only on the later years.

[11] The references to publications older than 1998 sum to 7,850,607 (out of 23,674,098) in the case of the *Science Citation Index*, and to 579,731 (out of 1,305,293) in the case of the *Social Science Citation Index*.

## Mapping of journal structures

One would not expect the two databases to be very different when the focus is on established fields of science and major journals. In order to test for this baseline, we used two major journals: the *Journal of the American Chemical Society* (*JACS*) and this journal itself—that is, the *Journal of the American Society for Information Science and Technology JASIST*—and compare the resulting visualizations using the parameters as specified above. Both these journals are major journals in their respective disciplines and rich sources of citations. *JACS* had an impact factor of 7.885 in 2007. One of us has analyzed the structure of this journal's citation environment in other studies (Leydesdorff, 1991; Leydesdorff & Bensman, 2006). *JASIST* had an impact factor of 1.436 in 2007. The resulting maps of the citation environments of this journal may be easier to compare (and validate) for the readership.

*a. The citation impact environment of the JACS.*

Figures 1 and 2 show the citation impact environments of the *JACS* in the *Scopus* database and the ISI database, respectively. In both cases, 18 journals are drawn into the analysis. However, the two lists of journals are not identical.

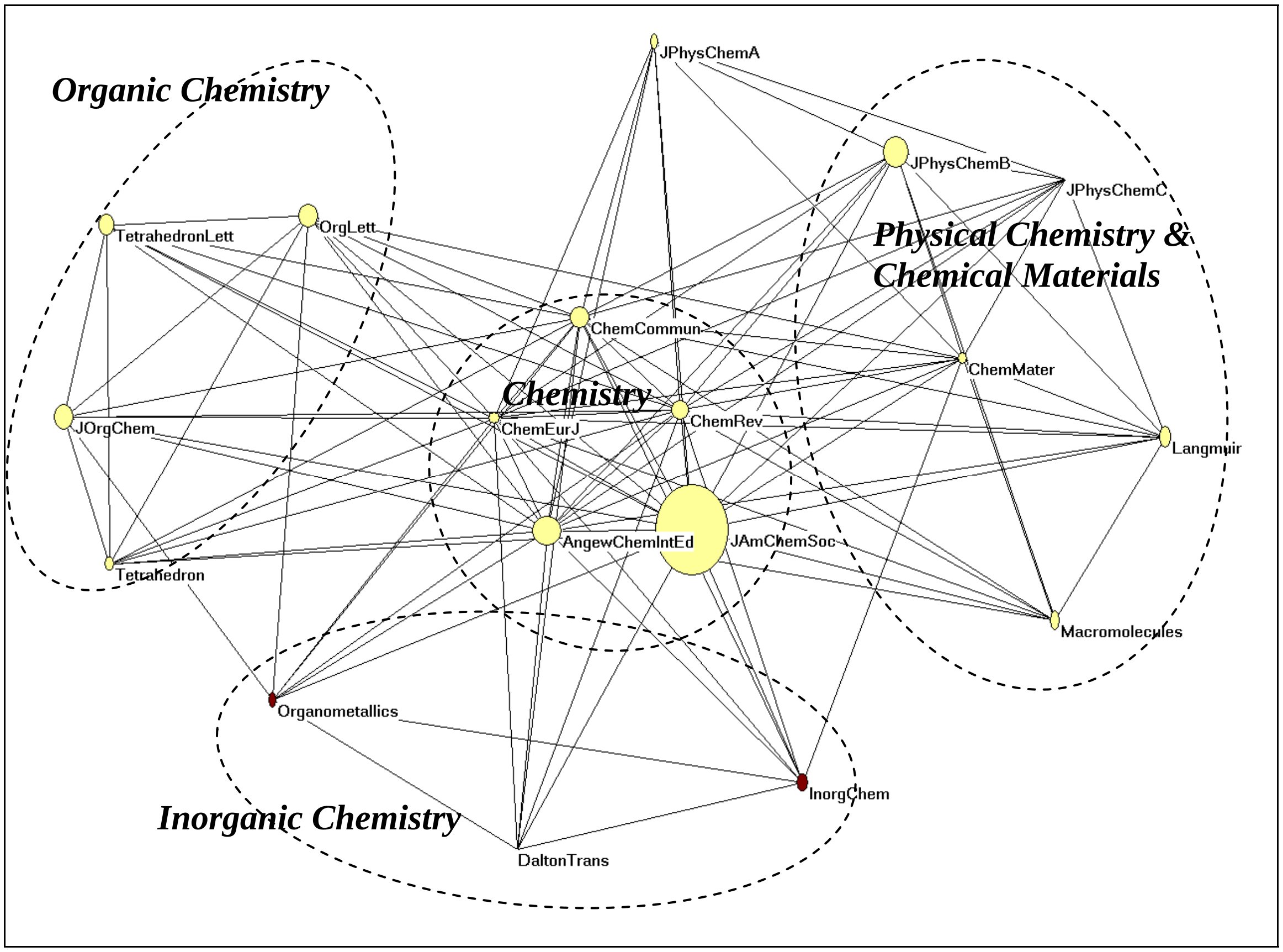

**Figure 1**: The citation impact environment of the *JACS* in *Scopus*; *N* = 18; threshold 1%; cosine > 0.2.

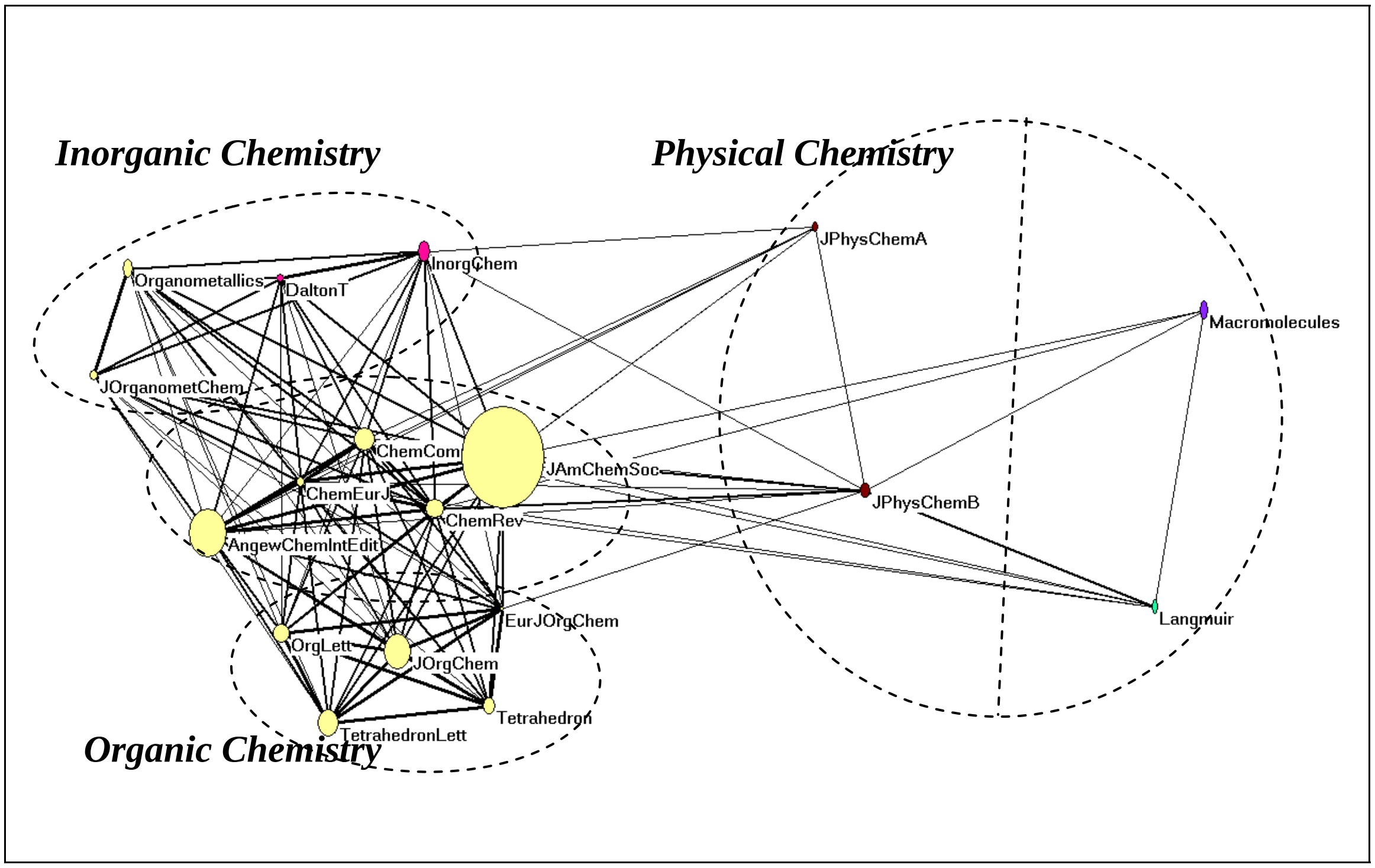

**Figure 2**: The citation impact environment of the *JACS* in the *SCI*; *N* = 18; threshold 1%; cosine > 0.2.

Some differences between the two maps are generated because journals are sometimes included in the one database but not in the other. For example, the *Journal of Physical Chemistry C* is not included in the ISI database; the *Journal of Organometallic Chemistry* does not pass the threshold (of 1,831 citations) in *Scopus* by contributing only 1,512 to the citations of the *JACS*.

In both Figures 1 and 2, the main structures can easily be recognized, but these representations differ significantly in the details. Factor analysis of both citation matrices provides a similar structure in terms of four factors explaining 72.9% of the variance in the case of *Scopus* and 79.1% in the case of the ISI. In the solution for the ISI database, *Langmuir* and *Macromolecules* are more clearly distinguished from the *Journal of*

*Physical Chemistry A and B,* while the case of *Scopus*, the *Journal of Physical Chemistry C* and *Chemical Materials* were drawn into this component.

Since other journals are included in both ego-networks, one percent of the total number of citations will also be different in the two databases. One would expect this number to be higher in the journal-journal citation data based on *Scopus* than in the one based on ISI database because more journals are included. However, the total number of citations to *JACS* is lower (183,186) in the journal-journal citation matrix based on the *Scopus* database, and the total number of references in *JACS* is 77,778, including 14,985 self-citations. These numbers are 295,465 for citations to *JACS* (+ 61.3%), 120,506 references from articles in *JACS* (+ 54.9%), and 20,706 self-citations (+ 38.2%) in the journal-journal citation data based on the ISI database (Table 3). The citation frequencies among journals are also different between the two journal-journal citation matrices. The network in Figure 1 is accordingly considerably less dense than in Figure 2.

| *JACS* | *Scopus* (1996-2007) | *SCI* (1998-2007) | *SCI* |
|---|---|---|---|
| Total cited in ego network | 183,186 | 73,414 | 295,465 |
| Total citing in ego network | 77,778 | 38,192 | 130,506 |
| Self-citations | 14,985 | 14,725 | 20,706 |

**Table 3**: comparison of cited and citing patterns of *JACS* in the three sets.

As noted, the ISI database allows us to select data only for the last ten years. The number of within journal self-citations for *JACS* decreases to 14,725 of the 20,706 citations in the unrestricted case (that is, 71.1%). This is very close to the 14,985 self-citations found in the *Scopus* database. Comparison among the columns of Table 3 positions the *Scopus* (1996-2007) set between the *SCI* set and the reduced *SCI* (1998-2007) as was to be expected.

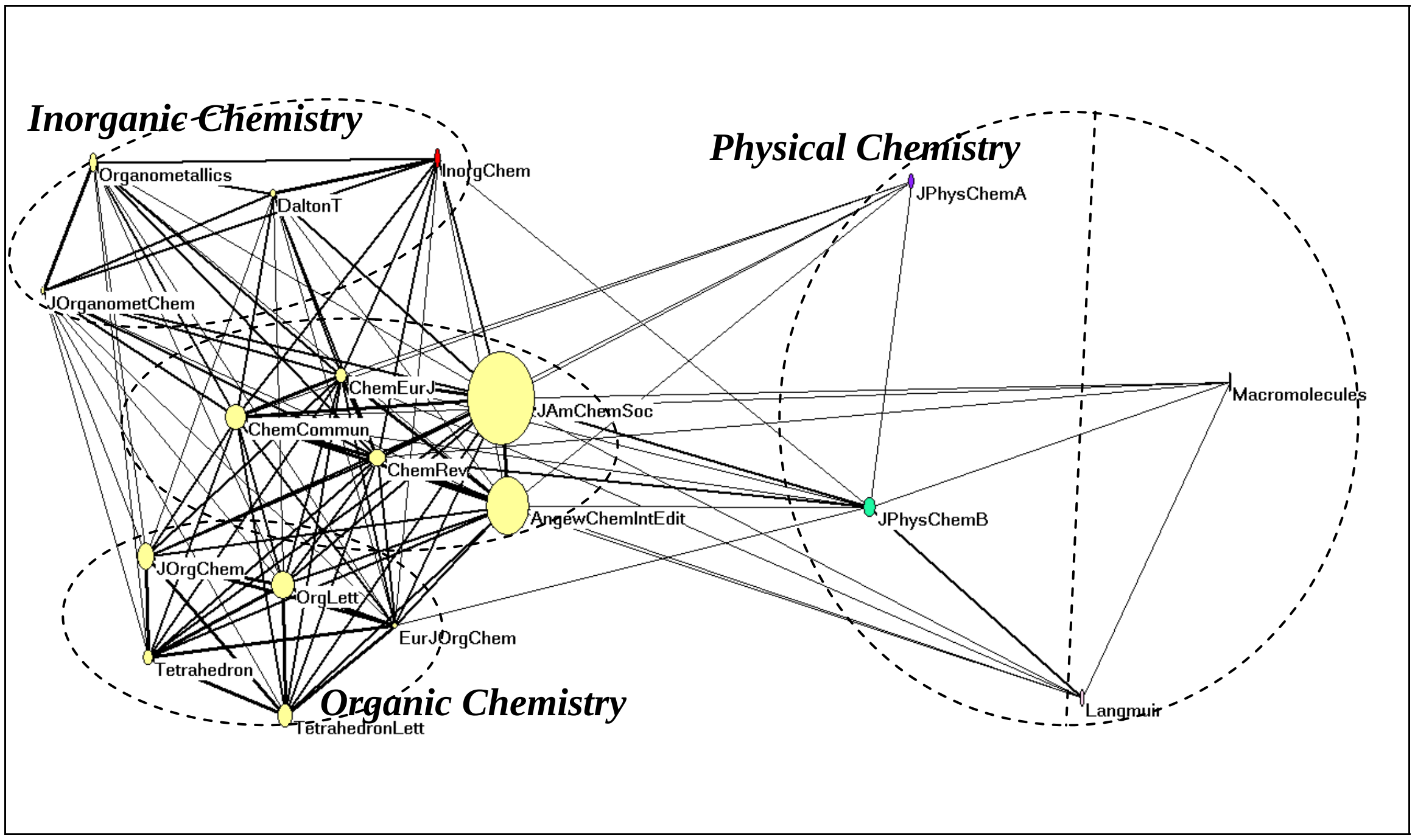

**Figure 3**: The citation impact environment of the *JACS* in the *SCI* (1998-2007); $N = 18$; cosine > 0.2.

If we apply the same set of 18 journals used in the full analysis to the ISI data for 1998-2007, the structure of the data remains virtually the same (Figure 3). The two citation matrices (with and without the references to publications older than ten years) correlate with $r = 0.978$ ($p < 0.001$). Structure is also maintained: the correlation between the four-factor matrices with and without the older data is $r = 0.993$ ($p < 0.001$). Thus, this ISI data is not structurally affected by the older data, although they provide a substantial part (29.8% in this case) of the citations. However, the organization of the factor structure is different in the case of using the *Scopus* database (Tables 4 and 5, respectively). Both the order of the factors and the rank-orders of the variable loadings within the components are different.

| | Component | | | |
|---|---|---|---|---|
| | 1 | 2 | 3 | 4 |
| Tetrahedron | **.938** | | | |
| Eur J Org Chem | **.925** | .179 | -.117 | |
| J Org Chem | **.922** | .192 | -.122 | |
| Tetrahedron Lett | **.916** | | | |
| Org Lett | **.891** | .302 | -.135 | |
| J Am Chem Soc | | **.898** | -.135 | |
| Chem Rev | .202 | **.879** | | .169 |
| Chem-Eur J | .150 | **.864** | .273 | |
| Angew Chem Int Edit | .137 | **.787** | | -.108 |
| Chem Commun | .154 | **.785** | .363 | -.199 |
| J Organomet Chem | | | **.832** | |
| Dalton T | -.377 | .238 | **.807** | |
| Organometallics | -.204 | | **.761** | |
| Inorg Chem | -.360 | .388 | **.582** | |
| J Phys Chem A | -.244 | | -.225 | **.892** |
| J Phys Chem B | -.514 | | -.595 | **.227** |
| Langmuir | -.441 | | -.543 | *-.342* |
| Macromolecules | -.351 | -.213 | -.369 | *-.461* |

Extraction Method: Principal Component Analysis.
Rotation Method: Varimax with Kaiser Normalization.
a Rotation converged in 6 iterations.

**Table 4**: Factor matrix for the citation patterns among 18 chemistry journals in the citation impact environment of *JACS* in 2007 using the *SCI* (1998-2007). Four factors explain 77.6% of the variance.

| | Component | | | |
|---|---|---|---|---|
| | 1 | 2 | 3 | 4 |
| Chem Rev | **.915** | .237 | -.103 | -.152 |
| J Am Chem Soc | **.876** | | .228 | -.187 |
| Chem Eur J | **.800** | | -.320 | |
| Chem Commun | **.782** | | -.189 | .468 |
| Angew Chem Int Ed | **.752** | .277 | | .126 |
| Tetrahedron | .148 | **.912** | -.235 | |
| Tetrahedron Lett | .114 | **.900** | -.236 | |
| J Org Chem | .271 | **.896** | -.235 | |
| Org Lett | .346 | **.872** | -.221 | |
| J Phys Chem B | | -.196 | **.853** | -.241 |
| J Phys Chem C | | -.141 | **.849** | -.117 |
| Chem Mater | -.125 | -.263 | **.807** | .148 |
| Langmuir | -.269 | -.137 | **.664** | |
| Macromolecules | -.406 | -.132 | **.122** | |
| J Phys Chem A | | -.237 | | -.877 |
| Dalton Trans | | -.290 | -.303 | **.500** |
| Inorg Chem | .486 | -.515 | -.231 | **.301** |
| Organometallics | .345 | -.377 | -.372 | **.118** |

Extraction Method: Principal Component Analysis.
Rotation Method: Varimax with Kaiser Normalization.
a Rotation converged in 6 iterations.

**Table 5**: Factor matrix for the citation patterns among 18 chemistry journals in the citation impact environment of *JACS* in 2007 using the *Scopus* dataset, 1996-2007. Four factors explain 72.9% of the variance.

Both the figures and the structural components are rather different among the two databases, while they are not significantly different between the *SCI* and the reduced set of *SCI* (1998-2007). The comparison with *Scopus* (1996-2007), however, does not allow us to draw any inferences about which one of the two representations to choose.

b. *JASIST*

In the context of this study, *JASIST* provides us with an interesting case because one would expect this journal to have suffered from its name change from *JASIS* to *JASIST* in 2001, and perhaps from the previous name change from *American Documentation* into *JASIS* in 1969. In the case of a name changes, the historical record may not be carried over to the new name (Bensman & Leydesdorff, 2009).

| | *Scopus* 1996-2007 | *ISI* 1998-2007 | *ISI* |
|---|---|---|---|
| Total cited | 2091 (*402*) | 2595 (*165*) | 3966[12] (*189*) |
| *JASIS* | 686 (*232*) | 0 | 0 |
| *Am Doc* | 0 | 0 | 65 (*9*) |
| Total citing | 2,843 (*575*) | 3,663 (*315*) | 5,798 (*385*) |
| Self-citations | 430 | 588 | 868 |

**Table 6**: Citations and self-citations of *JASIST* in the *Scopus* and ISI databases, respectively. Italic numbers between brackets indicate the number of unique journal-journal citation relations involved.

Table 6 shows first that the name change of *JASIS* into *JASIST* affects the *Scopus* data, but not the ISI data.[13] The older name of *American Documentation* can still be retrieved in the ISI database as a non-source item with 65 references in 2007. While the difference between the two totals for "cited" can be explained in terms of the cutoff in the *Scopus* database before 1996, the difference in "total citing" is huge. These citations are provided by authors of articles published in *JASIST* during 2007: 315 unique journal-journal relations contain 3,663 references in the *SCI* 1998-2007 set against 575 unique journal-journal relations which carry only 2,843 references in the *Scopus* 1996-2007 set. The lower number may partly be due to the relative large share of book reviews in the *JASIS* previously (Ove Kahler, *personal communication*, 28 August 2009).

[12] The ISI provides a number of 3,026 for total cited in 2007 (in both the *SCI* and the *Social SCI*). This would include the category "All others" and must be erroneous since Total Cites should be larger than the sum of the citations in the matrix (excluding the category "All others"). The separate sums for the *SCI* and the *Social SCI* are 1,927 and 2,039, respectively.

[13] As explained in the help files of the *JCR* (at http://admin-apps.isiknowledge.com/JCR/JCR?RQ=TITLE_CHANGES), the ISI data is not affected because the name change did not change this title's position alphabetically.

The number of self-citations may vary because of differences in the respective definitions of citable issues and the time spans under consideration. The number of self-citations in the ISI database during the last ten years was 588 (of the 868 self-citations). The number of citations in *JASIST* 2007 to *JASIS* in the *Scopus* database is 133. Thus, we have to compare the 588 self-citations during the period 1998-2007 in the ISI database against 566 (= 430 + 133) self-citations to the set of both *JASIST* and *JASIS* during the period 1996-2007 in the *Scopus* database. Given this match, the lower number of references ("citing") in the *Scopus* database, must find its origin in the differences between the journal sets in the relevant citation environments of *JASIST*.

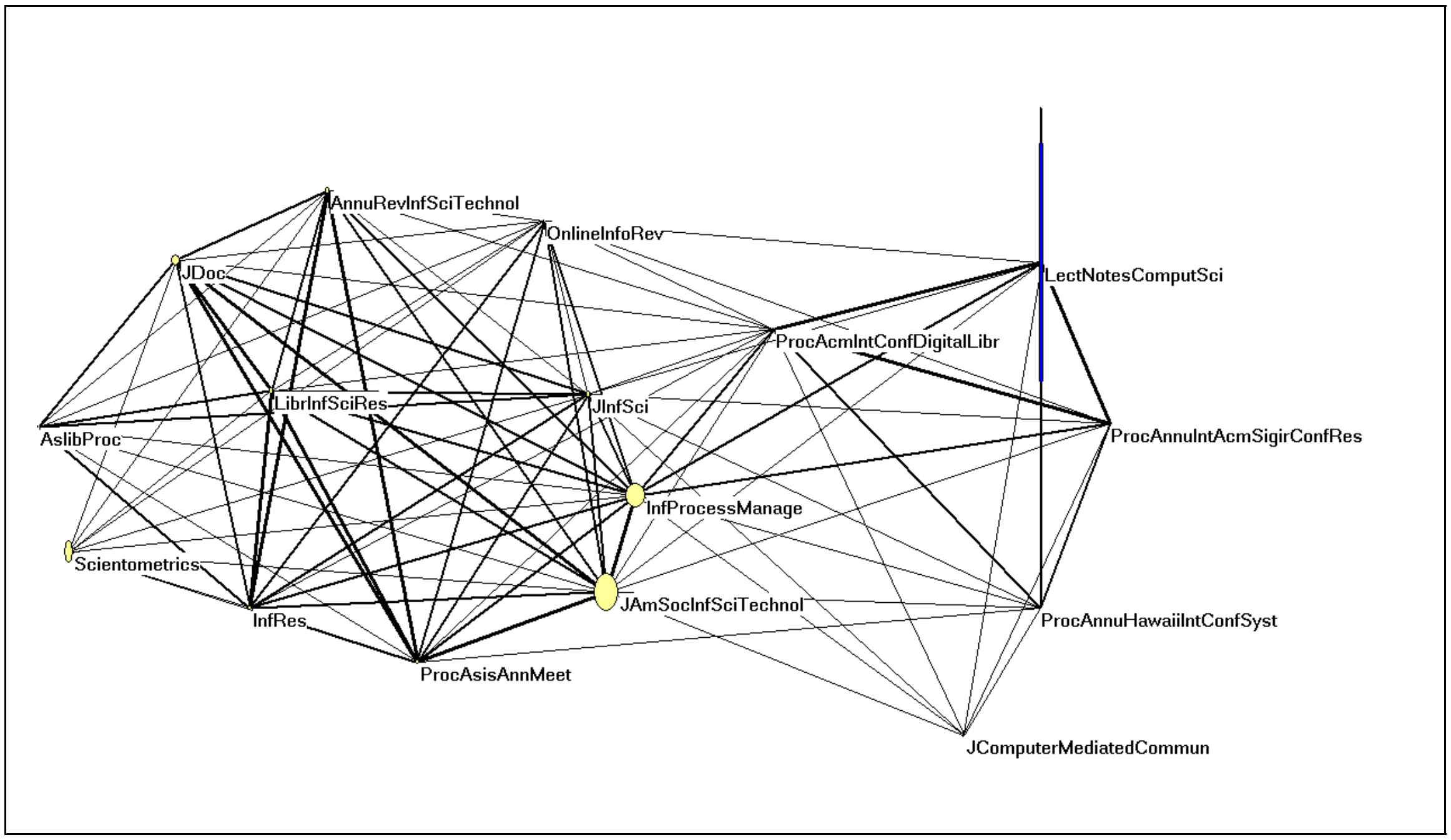

**Figure 4**: The citation impact environment of *JASIST* in the *Scopus* database; $N = 16$; threshold 1%; cosine > 0.2.

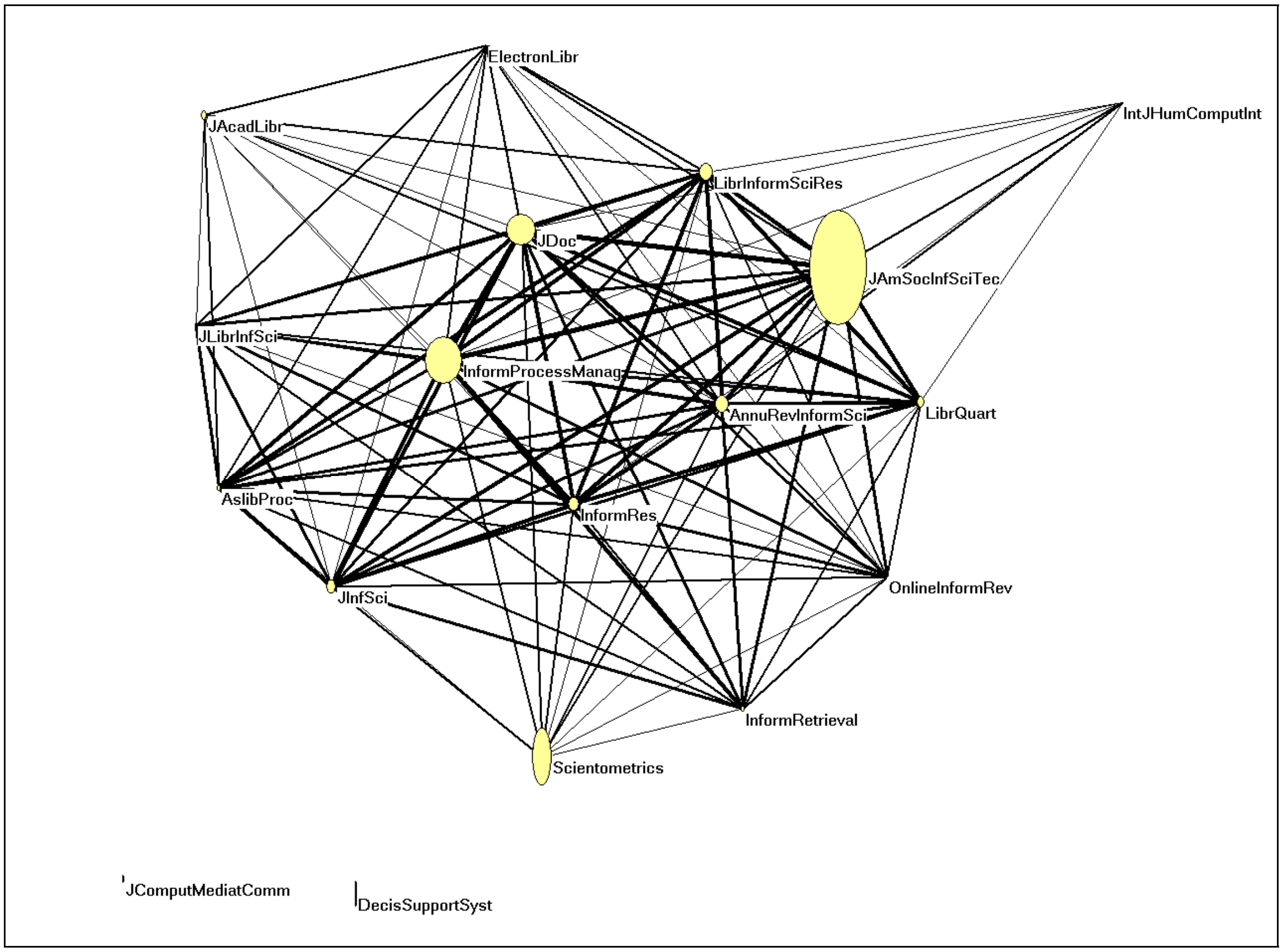

**Figure 5**: The citation impact environment of *JASIST* in the ISI database; $N = 18$; threshold 0.69%; cosine > 0.2.

The two maps using the full sets (Figures 4 and 5) are very different. Note that we have adjusted the threshold in the ISI case in proportion to the different numbers of total citations because of the name change in 2001 (in accordance with Table 6). With this threshold 18 journals are selected in the ISI environment as relevant. In the *Scopus* database, 16 journals are selected. These include a second set of computer science journals (Figure 4), while the ISI database provides us with a representation of library and information science as a relatively closed set of journals despite the lower threshold level (Figure 5).

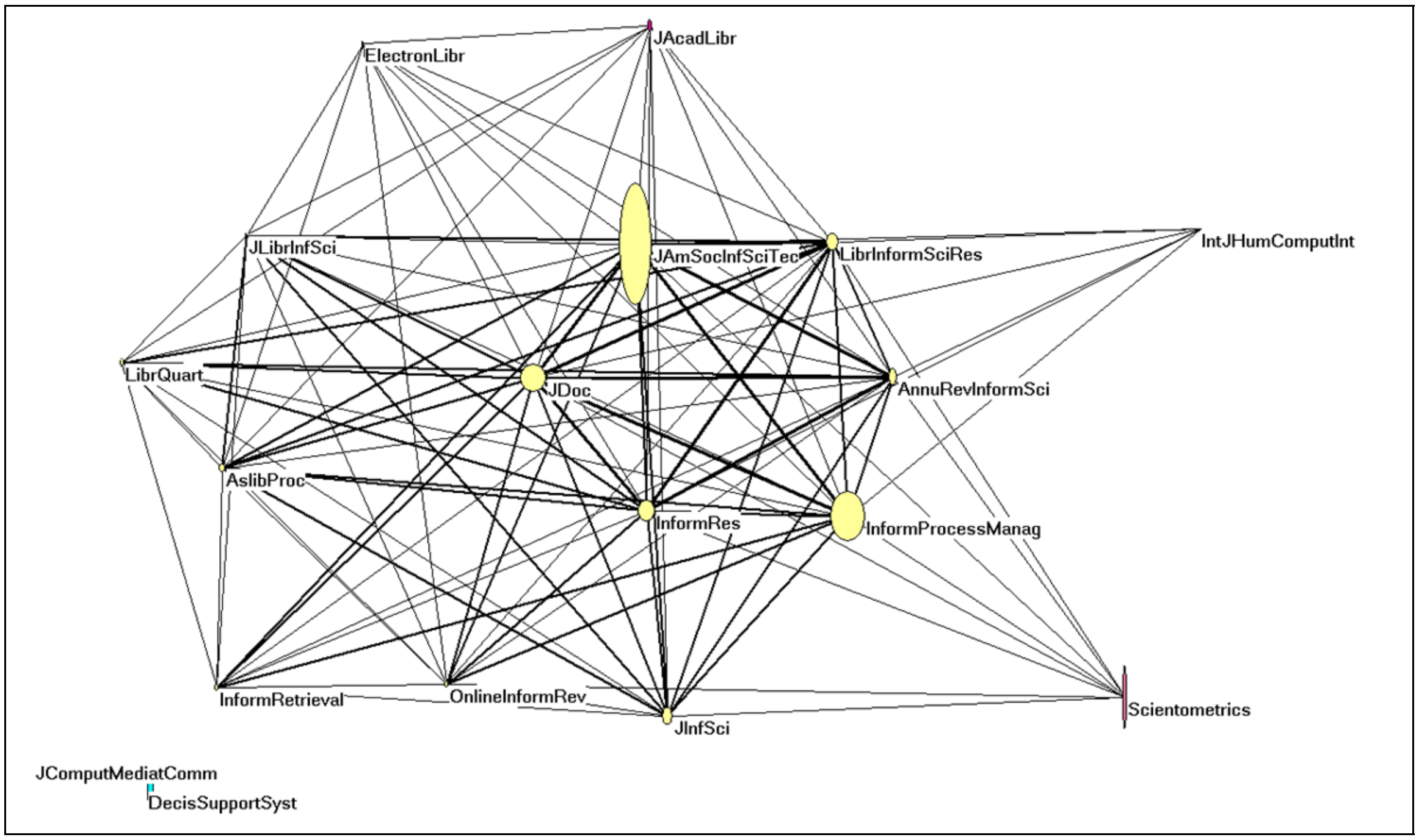

**Figure 6**: The citation impact environment of *JASIST* in the *SCI* 1998-2007 using the same set as in Figure 5 (including all citations).

We checked again against the matrix based on the same set of journals in the ISI database, but using citation data only for the last ten years (1998-2007). Although this reduces the grand sum of the matrix from 6,075 to 4.233 (69.7%), the two matrices are again virtually identical ($r = 0.995$; $p < 0.001$). Figure 6 shows the map for the *SCI* 1998-2007. The two figures are marginally different. For example, *Scientometrics* is more distanced from the set of information-science journals if older citations are not taken into account. As noted by Bensman & Leydesdorff (2009), the citation structure in the ISI database is not significantly affected by discarding the historical record. The citation structures are reproduced in this data yearly and can hence be considered as robust.

In summary, the networks in the cores of specialties are connected in the ISI database more strongly than in the journal-journal citation matrix based on the *Scopus* database. This is not because of the wider casting of the network in the latter case, but because of an erosion of structure as a consequence of lower average numbers of citations (cf. López-Illesca *et al*., 2008). Paradoxically, the maps from the *Scopus* database may appear more informative than the ones from the ISI database. The presence of more relevant environments provides more information in these representations. Using the ISI data, this same effect should be achievable by using a lower threshold. In other words, the ISI data is richer in structure than the *Scopus* data in this case. This is *not* caused by the historical record before 1996, which is available in the *JCRs*, but not in the *Scopus* (1996-2007) data. This historical record is large, but not important for the mapping.

**Extension to the humanities**

One obvious shortcoming of the ISI databases is the limitation of the *JCRs* to the sciences and the social sciences. The *Arts & Humanities Citation Index* (*A&HCI*) does not include a *JCR.* One of the promises of the *Scopus* database is the inclusion of journals from the humanities. As noted, the recent announcement of the *Scopus* team to increase the coverage of journals in the arts and humanities to 3,500 by adding approximately 1,450 top global journals using the European Reference Index for the Humanities (ERIH) of the European Science Foundation, is not covered by the current study which uses *Scopus* data until 2007.[3]

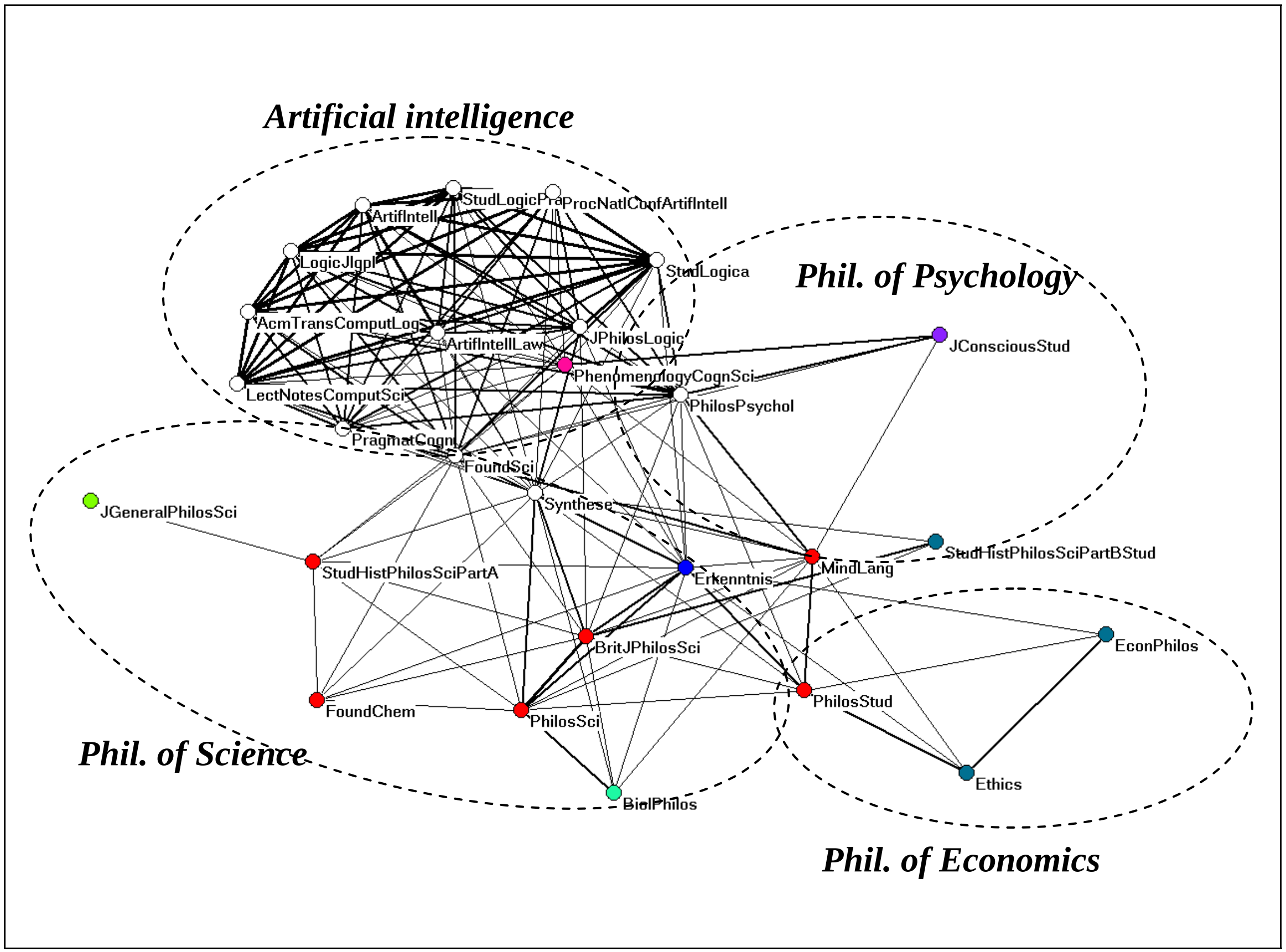

**Figure 7**: Citation impact environment of *Erkenntnis*; $N = 32$; threshold: 1%; cosine > 0.2.

Is the promise of more coverage than the ISI set in the social sciences and the arts & humanities fulfilled in the *Scopus* 1996-2007 set?[14] Using the journal *Erkenntnis* as the ego,[15] Figure 7 shows the citation impact environment of this journal using the same parameters as before. Thirty-two journals are included. The screeplot suggests a model with four or five factors. Table 7 shows the four-factor solution.

[14] For example: "Actually Scopus includes all of the Social Sciences titles in Thomson Scientific Social Sciences Citation Index®, as well as an additional few hundred titles." at http://info.scopus.com/detail/what/julie_arnheim.asp (retrieved on August 9, 2009).

[15] *Erkenntnis* was rated as an A-journal on both the provisional lists for philosophy and the history and philosophy of science of the European Reference Index for the Humanities at http://www.esf.org/research-areas/humanities/research-infrastructures-including-erih/erih-initial-lists.html.

**Rotated Component Matrix(a)**

| | Component | | | |
|---|---|---|---|---|
| | 1 | 2 | 3 | 4 |
| Acm Trans Comput Log | **.986** | | | |
| Logic J Igpl | **.982** | | | |
| Lect Notes Comput Sci | **.980** | | | |
| Artif Intell | **.979** | | | |
| Stud Logic Pract Reasoning | **.979** | | .100 | |
| Stud Logica | **.953** | | | |
| Proc Natl Conf Artif Intell | **.946** | | | |
| Artif Intell Law | **.828** | | | |
| J Philos Logic | **.789** | .138 | -.117 | |
| Pragmat Cogn | **.554** | | .283 | |
| Found Sci | **.529** | | .239 | -.356 |
| Philos Sci | -.102 | **.871** | -.124 | -.173 |
| Brit J Philos Sci | | **.860** | -.202 | -.174 |
| Synthese | .241 | **.769** | | |
| Erkenntnis | | **.689** | | .188 |
| Biol Philos | | **.454** | | -.189 |
| Stud Hist Philos Sci Part B Stud | | **.359** | -.258 | -.266 |
| J Radiol | | -.186 | -.128 | |
| European Review | | -.183 | -.122 | |
| Zool Scr | | -.171 | -.130 | |
| J Conscious Stud | | -.103 | **.755** | -.132 |
| Phenomenology Cogn Sci | .115 | -.131 | **.748** | -.241 |
| Philos Psychol | .463 | .144 | **.737** | |
| Mind Lang | | .261 | **.612** | .349 |
| J Hist Biol | | -.116 | -.123 | -.107 |
| Ethics | -.132 | | | **.750** |
| Econ Philos | | | -.133 | **.593** |
| Philos Stud | | .473 | .229 | **.585** |
| Stud Hist Philos Sci Part A | -.103 | .142 | .155 | *-.386* |
| Found Chem | | .199 | | *-.300* |
| J General Philos Sci | | | | *-.248* |

Extraction Method: Principal Component Analysis.
Rotation Method: Varimax with Kaiser Normalization.
a Rotation converged in 12 iterations.

**Table 7**: Four-factor model of the citation impact environment in the ego-network of *Erkenntnis* in the *Scopus* database, 2007.

Four factors explain 54.8% of the variance. The citation pattern of *Erkenntnis* itself loads on a second factor with five other journals in the philosophy of science. These five other journals are all included in the *Social Science Citation Index*. However, three more journals included in the *Scopus* database can also be categorized as philosophy of science (*Studies in History and Philosophy of Science Part A*, *Foundations of Chemistry*, and the *Journal for General Philosophy of Science*). These three journals have in common that they load with a negative sign on factor 4 in Table 7.

Unfortunately, the *Scopus* (1996-2007) database did not otherwise contain the rich set of humanities journals that are included in the *Arts and Humanities Citation Index* of the ISI. Three journals, for example, had the word "art" in their title: the *International Journal of Art and Design Education*, the *American Journal of Art Therapy*, and *Art Therapy*. The first of these three journals is included in the *Social Science Citation Index*; the latter two are not. However, they cannot be considered as art journals.

A large set of journals (more than 100) containing the word "art" in their titles are included in the *A&HCI*. *Leonardo*, for example, can be considered as a leading journal for readers interested in the applications of contemporary science and technology to the arts (Salah & Salah, 2008). Despite its relative ranking as a C-journal on the lists of the European Research Index for the Humanities, the journal is included in the *A&HCI* but not yet in the *Scopus* database at the time of this investigation.[16] The inclusion in the *A&HCI* provides us with references in this journal from the citing side. The citations to the journal in the ISI domain can also be retrieved.

[16] In the meantime (August 2009), *Leonardo* is covered by *Scopus* for the years 2008 and 2009, and partly for 2004.

It is less well-known that the all ISI-indexes (including the *A&HCI*) allow us to construct a journal citation map from the cited references in the articles aggregated into journals. The cited reference field is highly standardized across the three ISI-*Indices.* For example, Figure 8 can be constructed using the 3,259 references in the 107 articles which cited *Leonardo* in 2008. The aggregate of their reference lists can be used for a "cited" journal map. Similarly, one can use all references in the 157 articles published in *Leonardo* in 2008 for a "citing" journal map. [17]

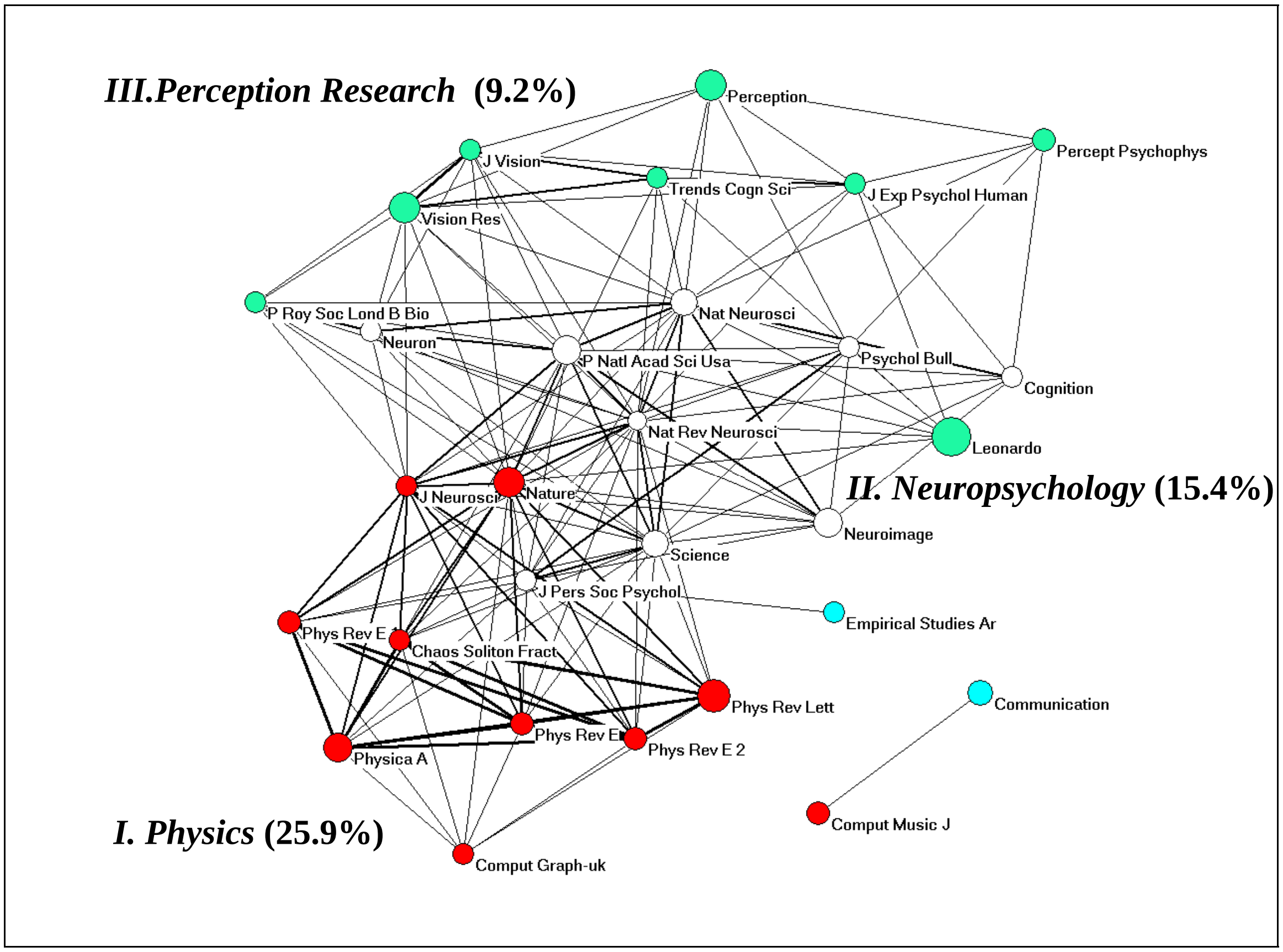

**Figure 8**: Cosine relations among 3,259 references in 107 articles citing *Leonardo* during 2008; threshold 0.5%; cosine > 0.2.

[17] Figure 7 was made using BibJourn.Exe which is freely available for non-commercial usage at http://www.leydesdorff.net/software/bibjourn/index.htm .

Three factors in this case explain 50.3% of the variance. (Since the organization of the factors does not map in terms of contingent areas in the cosine-base map, we used colors to indicate the three factors in this case.) The self-citation pattern of articles in *Leonardo* citing *Leonardo* is attributed to the third factor (designated by us as "perception research"), but with a factor loading as low as 0.165. Its loading on the "neuro" factor is negative.

Using this methodology, one can make these maps for all journals in the three citation indices of the ISI. Since the structure of the reference patterns in the *Scopus* database is more complex than the uniform format in the ISI data, it hitherto is not possible to use *Scopus* data for this mapping by aggregating documents. The *Scopus* data requires some formatting for an unequivocal translation into the ISI format as used by *HistCite*™ and other programs.[18]

In summary, the *Scopus* database enriches our possibilities to map the arts and humanities because of its larger journal set. However, the possibilities to map also journals in the *Arts & Humanities Citation Index* of the ISI are larger than hitherto exploited. The coverage of the arts and humanities in the *A&HCI* was at the time of this research considerably larger than in the *Scopus* database.

[18] A number of these programs are available as freeware at http://www.leydesdorff.net/indicators/index.htm . One of these programs is Scop2ISI.Exe which enables the user to reformat downloads from *Scopus* into the tagged format of the ISI data.

## Extension to interdisciplinary journals

The extension of the journal set in the *Scopus* database allows us to address the position of peripheral journals which could previously not be addressed. Recently, one of us used citations to non-source journals in the ISI database to map the position of *Science & Public Policy* (Leydesdorff, 2008, p. 283). The *Scopus* database, however, includes this journal (Figure 9).

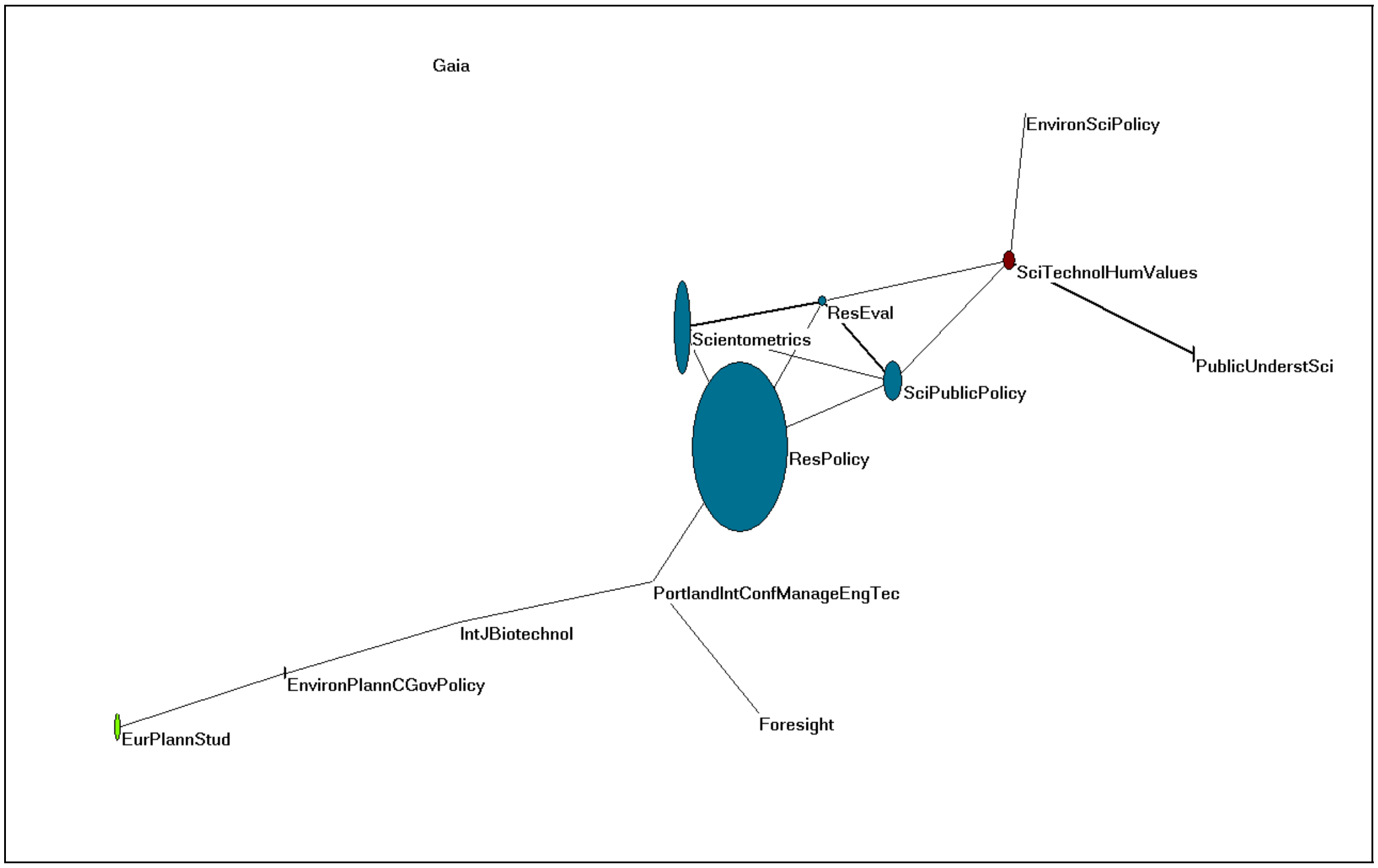

**Figure 9**: The citation impact environment of *Science & Public Policy* in the *Scopus* database in 2007; $N = 13$; threshold: 1%; cosine > 0.2.

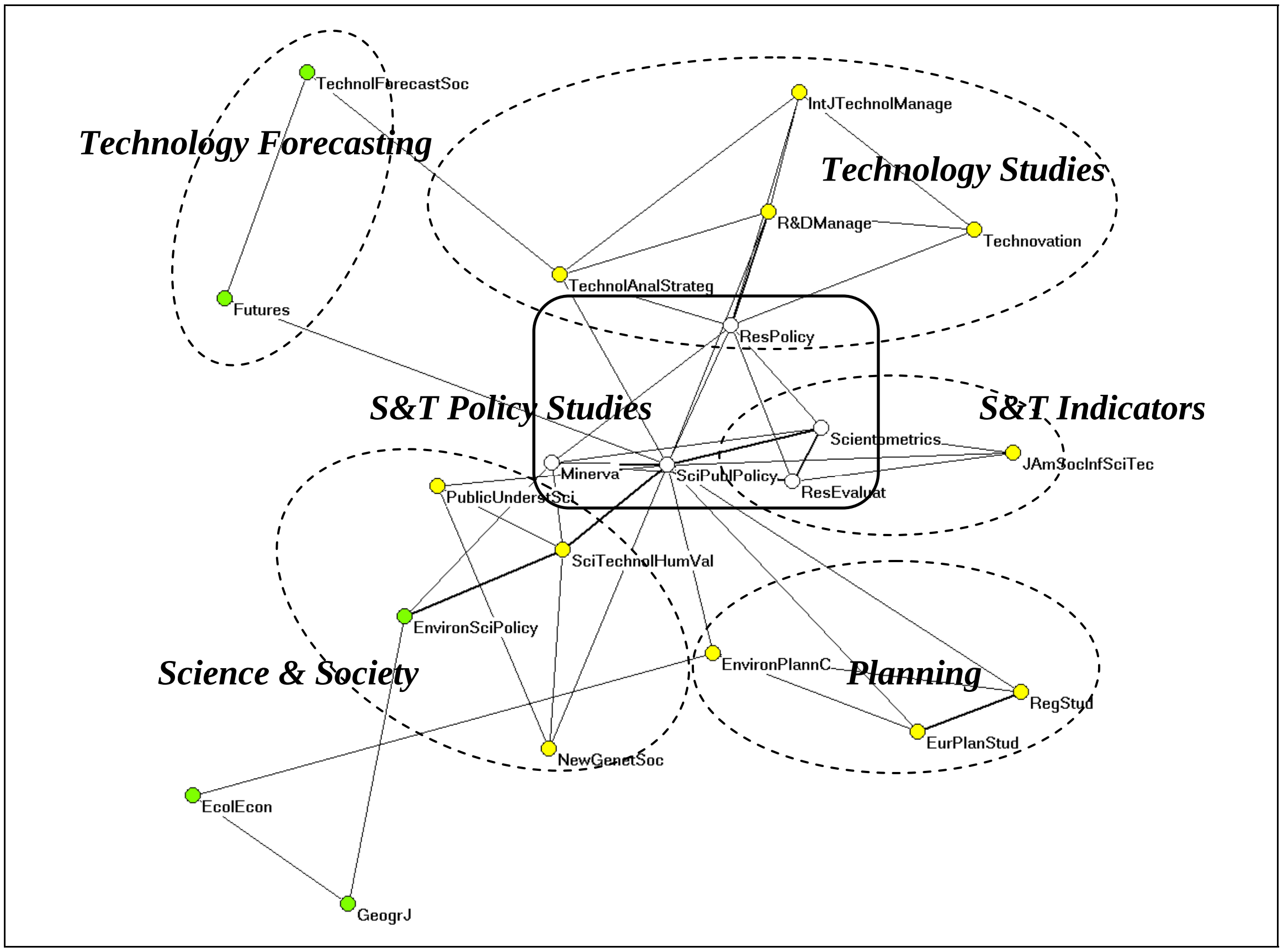

**Figure 10**: The citation impact environment of *Science & Public Policy* in the *Social Science Citation Index* (1998-2007); *N* = 21;[19] threshold: 1%; cosine > 0.2.

In order to pursue the comparison, let us add in this case the citation data about *Science & Public Policy* as a single non-source journal to the *Social SCI* (1998-2007). Figure 9 shows an organization on the basis of the *Scopus* (1996-2007) database that is very different from the one possible on the basis of the data in the *Social SCI* (1998-2007) in Figure 10.

In both cases, *Science & Public Policy* connects a number of journals in science and technology studies. The *Scopus* database shows a relatively heterogeneous set connected in a core group, while Figure 10 provides a more differentiated map of science and technology studies. Four factors explain 30.1% of the variance in this case,[20] while in the *Scopus* case the configuration is to a large extent (18.0%) defined by the first factor containing the core group which is indicated in both sets. This core group shows factorial complexity—sometimes used as an indicator of interdisciplinarity (Leydesdorff & Cozzens, 1993; Van den Besselaar & Heimeriks, 2001)—in the ISI set.

[19] *Science & Public Policy* itself has to be added to the list using a threshold, because it is not cited by itself in this representation since it is not processed by the ISI as a source journal.

The function of the interdisciplinary journal connecting different parts of this field of studies can be made visible in both representations. In previous studies, the interdisciplinary field of science and technology studies was found to be so weakly connected that it could no longer be mapped as integrated using the *JCRs* of the (*Social*) *Science Citation Index* (Leydesdorff, 2007; Leydesdorff & Van den Besselaar, 1997; Van den Besselaar, 2000). The addition of interdisciplinary journals thus shows the cross-connections as relatively weak ties with potentially important functions. One is more likely to find these journals included in the *Scopus* set than in the *SCI*, but the latter database allowed in this case for a richer representation by including the interdisciplinary journal from the list of non-source items.

**National subfields**

The claim is often made that the ISI databases are not suited for the evaluation of the social sciences and the humanities because these scholarly discourses are embedded in national cultures with a specific character. Hagendijk & Smeenk (1989) introduced the terminology of "national subfields" for field sciences like ecology, but the point has mainly been made more emphatically with reference to the specific character of continental European sociology and philosophy, where the premises are different from those in American and English literature (Nederhof *et al.*, 1989). The *Scopus* database promises to enable us to study these national sets in more detail using leading journals in the respective cultures. Let us focus on sociology in Germany and France as relevant examples (Glänzel & Schöpflin, 1999; Yitzhaki, 1998). Because of the potentially low numbers of citations in this field we included the complete citation impact environments—that is, without using a threshold—in these cases.

[20] The corresponding value is 31.6% in the case of the full set including the tails.

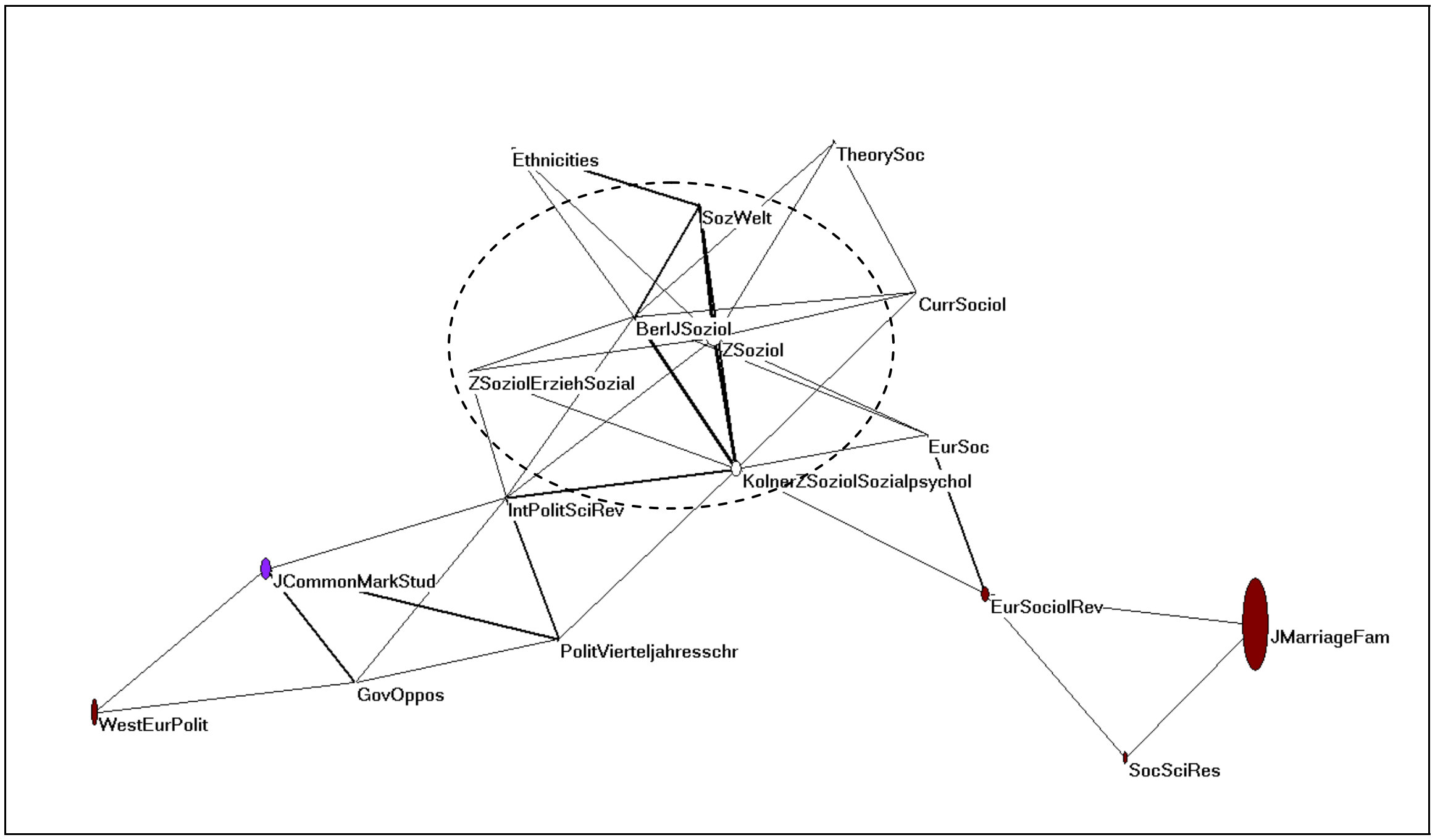

**Figure 11**: The ego-network of the *Zeitschrift für Soziologie* in the *Scopus* database; $N$ = 17; no threshold; cosine > 0.2.

Figure 11 provides the representation of the ego-network of the German *Zeitschrift für Soziologie* in the *Scopus* (1996-2007) domain. The journal is visible in its national context of a core group of journals in the social sciences in the German language. The international dimension is connected via the political-science journals involved in the German network, on the one side, and mainly via the European dimension, on the other. There are no citation relations to the leading American journals in this ("being cited") direction.

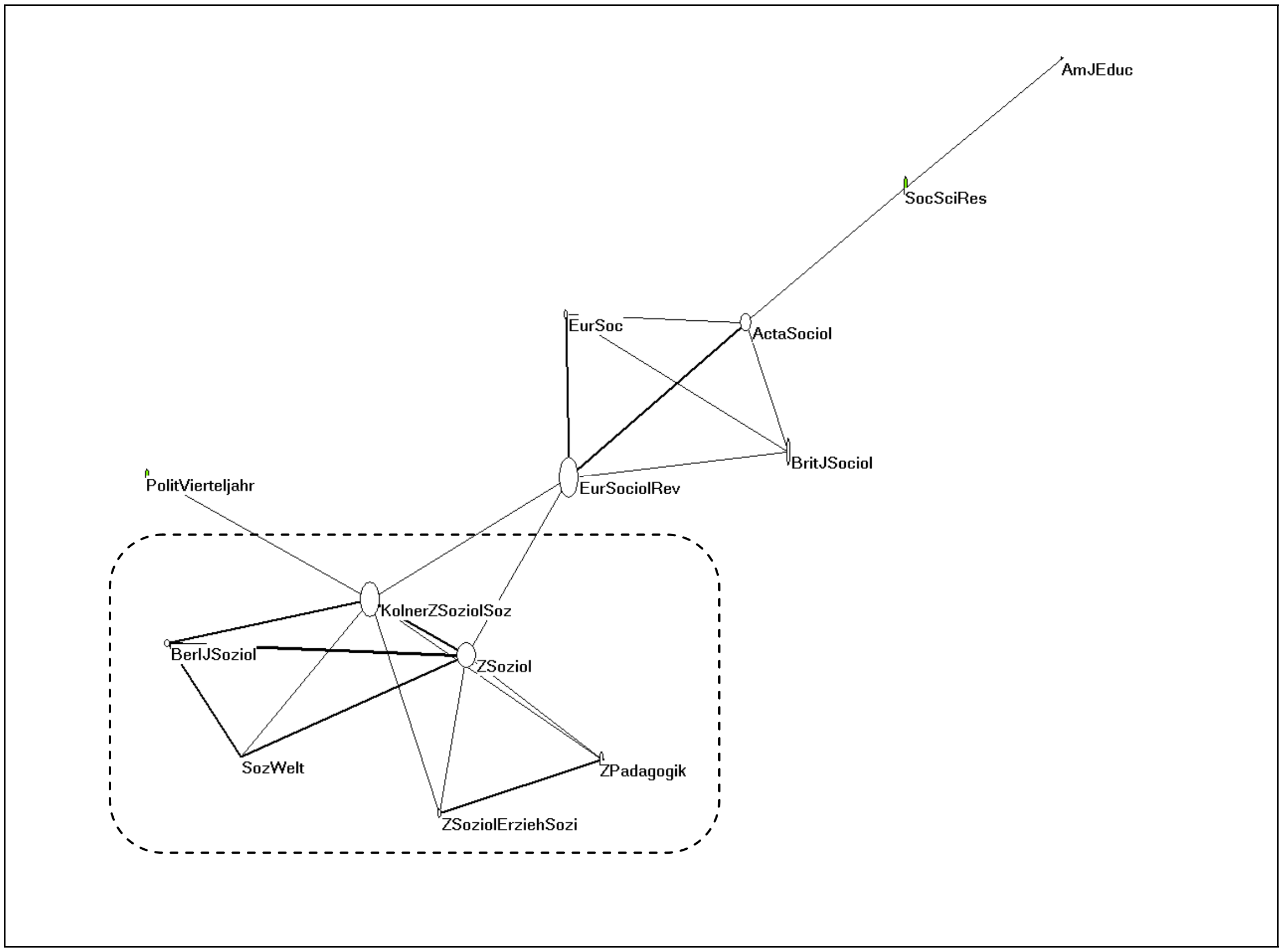

**Figure 12**: The citation impact environment of the *Zeitschrift für Soziologie* in the *Social Science Citation Index* 1998-2007; *N* = 13; no threshold; cosine > 0.2.

The *Zeitschrift für Soziologie* is also included in the *Social Science Citation Index.* Figure 12 provides the map of its ego-network in this database when the historical tail is removed (*SCI* 1998-2007). Adding the tail makes only a single additional relation visible between the *British Journal of Sociology* and the *British Journal of Sociology of Education,* and the network then is a bit denser. Although rather different again, the map based on the *Scopus* data seems to outperform the map based on the *SCI* even if the latter is not restricted to the 1998-2007 period.

In France, the national tradition in sociology is more integrated with the Anglosaxon literature than in Germany. Several (e.g., Canadian) journals accept manuscripts in both languages. A leading journal in French sociology—initiated by Pierre Bourdieu in 1975—is *Actes de Recherche en Sciences Sociales.* This journal belongs to the group of journals not cited in the *Scopus* database, but it is included in the *Social Science Citation Index* 2007. According to the *Scopus* database, *Actes de Recherche en Sciences Sociales* contains seven references during 2007 as against 766 references according to the *Social Science Citation Index* for that same year. *Actes de Recherche en Sciences Sociales*, however, had an impact factor of 0.074 in 2007.

A journal in French sociology which is included in both databases is the *Revue Française de Sociologie* with an impact factor 2007 of 0.222. We used this journal for the comparison. Figure 13 provides the map for the *Scopus* 1996-2007 data and Figure 14 for the complete set in the *Social Science Citation Index.* The restricted set in this case provides a map even poorer than Figure 14 (based on the full *SCI* data), while the latter is still outperformed by the one based on the *Scopus* (1996-2007) database in Figure 13.

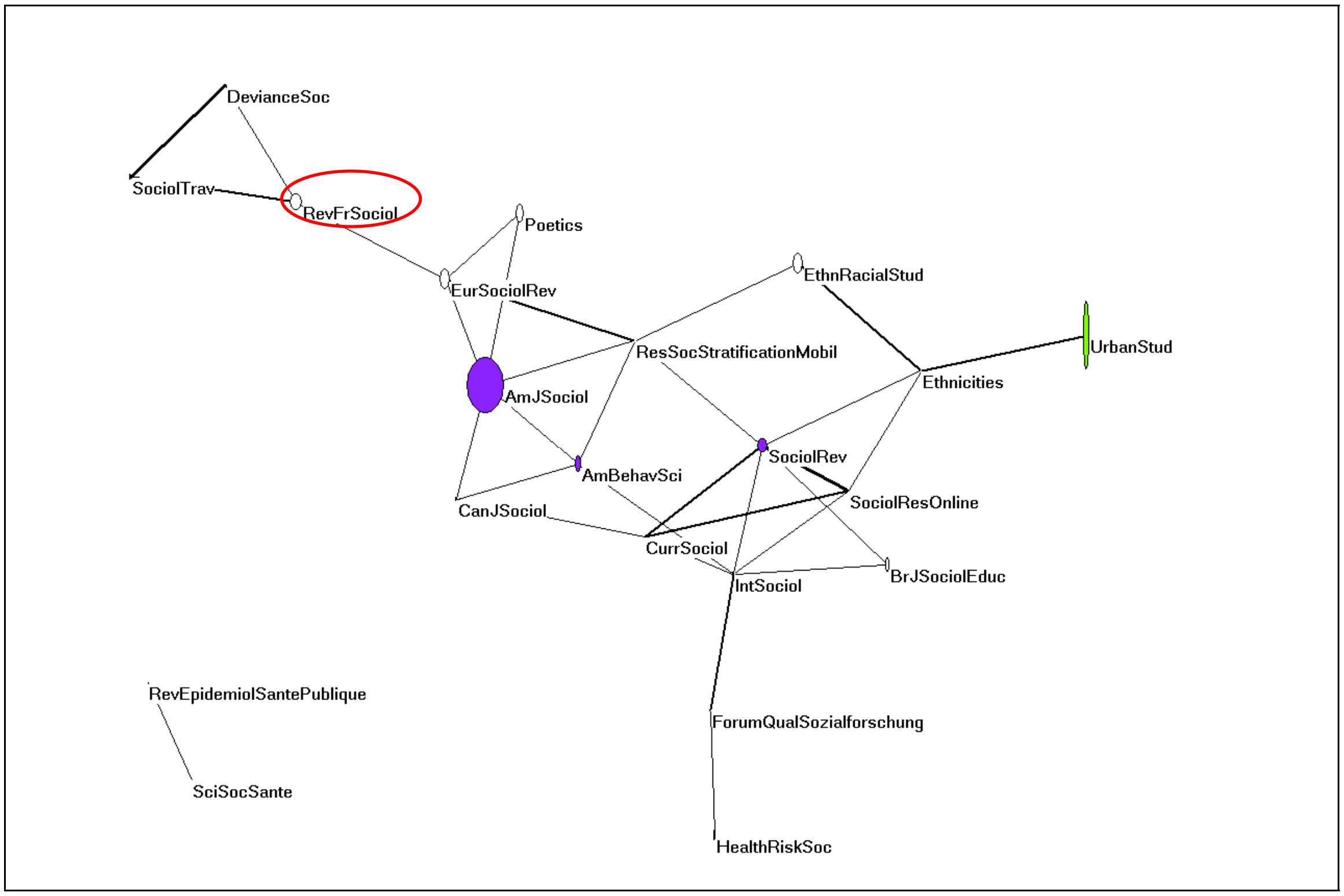

**Figure 13**: Complete citation impact environment of the *Revue Française de Sociologie* in the *Scopus* database 1996-2007 ; no threshold ; cosine > 0.2.

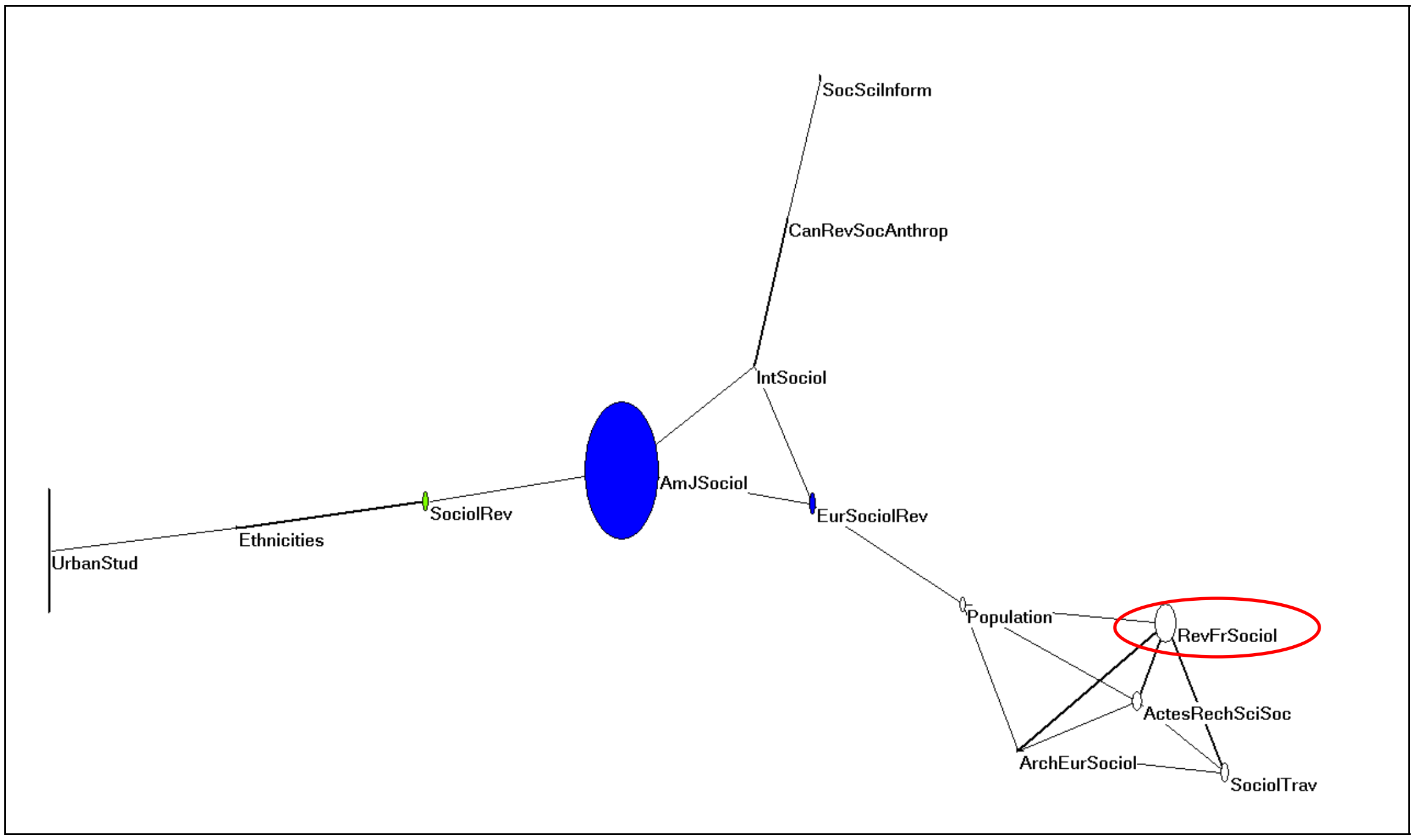

**Figure 14**: Complete citation impact environment of the *Revue Française de Sociologie* in the *Social Science Citation Index* (all years); no threshold ; cosine > 0.2.

Both figures show that the French sociological tradition is less distanced from the Anglosaxon discourse than the German one. In both the French and German cases, the maps made on the basis of the *Scopus* data outperform the ones on the basis of the *Social Science Citation Index.* Because of this focus on the social sciences, one would perhaps expect the historical tails of the distributions (in the *Social Science Citation Index*) to be more influential than they are. In other words, the loss of the tail in the *Scopus* database is more than compensated by the wider reach of this database.

**Non-scholarly journals and newspapers**

The *New York Times*, *Washington Post*, and the *Wall Street Journal* can be analyzed in terms of their citation patterns in the ISI database using the non-ISI source references. However, the journal names and abbreviations are not standardized in this list. In the *Scopus* database, these names are standardized, but as in the case of the ISI the newspapers are not processed in terms of their citing references but registered only when they are cited in the *Scopus* domain.

There are only few citations to newspapers in the matrix based on the *Scopus* database (e.g., 43 to the *NY Times* and 14 to the *Wall Street Journal*), because newspapers are not among the source journals. In contrast, the *Wall Street Journal* was cited 2,196 times by 354 source journals of the ISI databases in 2007. (This newspaper is mostly cited in law review journals.) However, neither database allows for systematic research at the interfaces between scholarly communication and newspapers. Citation relations are typical for the scientific domain; co-word analysis among titles or full texts may be a better tool if one is interested in analyzing developments at the interfaces of scientific communication (Callon *et al.*, 1986; Leydesdorff & Hellsten, 2005).

## Conclusions and discussion

By providing aggregated journal-journal citation data, both *Scopus* and the ISI databases enable us to map journals and to delineate specialties, while this cannot easily be done using *Google Scholar*. As is well known, the coverage of journals by *Scopus* is larger than by the ISI. Since the core groups of scholarly journals are to a large extent the same, *Scopus* can be expected to provide mainly an enlargement when compared with the ISI set. One can expect that the added journals are more peripheral, and that the added networks will thus be less dense than those among core journals (López-Illescas *et al.*, 2008).

In addition to this effect, the journal-journal citation matrix based on the *Scopus* data contains sometimes fewer citations to and references from individual journals than the one based on ISI data. Although the aggregations under the category “All others” have a considerable effect in the latter database, even after correction for this category, the numbers in the ISI database tend to remain higher than in the *Scopus* (1996-2007) database. The matrix based on *Scopus* is not only sparser but also less pronounced in terms of the values in the cells. Although this may significantly affect the analysis, a user may feel well served with a map showing a less pronounced, but wider net.

One major difference between the two databases is the inclusion of citations from the historical record in the ISI database, while the *Scopus* database begins only in 1996. We noted that this historical record is of the order of more than 30%, although this can be expected to vary among fields of science. In the case of the ISI database, the historical record is also reorganized in the event of a name change of a journal, but this affects the structure present in the database only marginally (Bensman & Leydesdorff, 2009). Otherwise, the citation structure is continuously reproduced and updated by the *JCRs*. The differences with the matrix based on the *Scopus* database find their origin in other rules for the inclusion of citations during the period covered by the database.

The extension of *Scopus* to include journals in the humanities could provide us with a new opportunity as long as a *JCR* for the *A&HCI* is missing. Figure 7 provided a journal map for a philosophical domain (using *Erkenntnis* as a source journal) which one cannot make with the ISI databases. Similarly, a large number of journals are processed by *Scopus* which are not processed by the ISI. We took *Science & Public Policy* as an example. Since these journals relate in relatively sparse parts of the network, inclusion of their relatively "weak ties" may provide new perspectives on the data (Granovetter, 1982).

In the case of *Science & Public Policy*, for example, a specific coherence among journals in the area of Science & Technology Studies can be revealed which cannot so easily be found using the ISI set. *Scientometrics*, for example, is firmly embedded in another cluster of journals—that is, library and information science—in the latter database. However, a relevant map could also be provided about the journal structures in this case using the references to non-source items in the ISI data. It seems worth the additional effort to use this ISI data in such cases. With some additional effort, journals in the arts and humanities can also be mapped in terms of aggregated journal-journal citations directly from the *Arts & Humanities Citation Index*.

Within the social sciences, it is common to complain about the inaptness of the American-based citation indices for the evaluation of nationally oriented cultural analyses such as those published in sociology journals in the respective languages. Does the wider coverage of *Scopus* solve this problem? In the case of leading journals in German sociology, we found the two databases to point in different directions, but both indicated a prevailing isolation of this tradition in terms of its being-cited patterns. The French tradition is less isolated and better perceived in main-stream sociological research, notably in the UK. This becomes better visible using the *Scopus* data than the *SCI* data.

Let us repeat that both databases do a fine job in providing source materials for the mapping of science. Some of the differences that are being shown in this paper are a result of the differences in maturity of the two databases. The ISI data are more mature in that the ISI has had decades longer to develop cleaning, standardization, and normalization procedures that show up in the raw data and JCR. The Scopus data are newer, and thus less clean (despite unique item identifiers) and less standardized in the cited references. However, the competition between the two may lead to further improvements such as the recent extension of the ISI database with 700 regional journals. Improvements in the *Scopus* database could focus on covering the source materials more systematically. As noted, a further extension of the coverage of journals in the arts and humanities was recently realized.

**Acknowledgement**

We are grateful for Almila Salah for her advice about using the *Arts & Humanities Citation Index* for the purpose of journal mapping. We are also grateful to an anonymous referee who provided us with additional information about the *Scopus* database.